\documentclass{ewic}
\usepackage{graphicx}
\usepackage[indentfirst=false,leftmargin=10pt,rightmargin=10pt]{quoting}

\begin{document}

\runningheads{Williams $\bullet$ Yao $\bullet$ Nurse}{Augmented Reality Tourism Apps through User-Centred Design}
\conference{Proceedings of British HCI 2017 - Digital Make Believe, Sunderland, UK.}

\title{\textit{ToARist}: An Augmented Reality Tourism App created through User-Centred Design}

\authorone{Meredydd Williams\\
Department of Computer Science\\
University of Oxford, Oxford, UK\\
\email{meredydd.williams@cs.ox.ac.uk}}

\authortwo{Kelvin K. K. Yao\\
Department of Computer Science\\
University of Oxford, Oxford, UK\\
\email{kelvin.khoo@stx.ox.ac.uk}}

\authorthree{Jason R. C. Nurse\\
Department of Computer Science\\
University of Oxford, Oxford, UK\\
\email{jason.nurse@cs.ox.ac.uk}}

\begin{abstract}
Through Augmented Reality (AR), virtual graphics can transform the physical world. This offers benefits to mobile tourism, where points of interest (POIs) can be annotated on a smartphone screen. Although several of these applications exist, usability issues can discourage adoption. User-centred design (UCD) solicits frequent feedback, often contributing to usable products. While AR mock-ups have been constructed through UCD, we develop a novel and functional tourism app. We solicit requirements through a synthesis of domain analysis, tourist observation and semi-structured interviews. Through four rounds of iterative development, users test and refine the app. The final product, dubbed \textit{ToARist}, is evaluated by 20 participants, who engage in a tourism task around a UK city. Users regard the system as usable, but find technical issues can disrupt AR. We finish by reflecting on our design and critiquing the challenges of a strict user-centred methodology.
\end{abstract}


\maketitle

\section{Introduction}

Augmented Reality (AR) projects virtual graphics into real environments, helping users absorb information in an intuitive manner. Such tools can be useful for navigation, enabling tourists to eschew paper-based maps. While AR once required head-mounted displays, smartphones can now support the technology. With mobile devices pervading our lives, tourism apps have grown in popularity. Although AR offers many advantages, tools have been criticised for usability issues. Displays are often cluttered with icons \citep{Julier2000}, a particular issue when overlays collide. Furthermore, AR apps have a limited field of view \citep{Tokusho2009} and often require an awkward stance. 

Usability is a key goal of user-centred design (UCD), where feedback is sought throughout the development process \citep{Vredenburg2002}. Through iteratively refining prototypes, the product is often better-suited to users' needs.

In this paper, we address usability in the context of Augmented Reality. Our contribution stems from the application of UCD to an AR tourism app, dubbed \textit{ToARist}. We first extract our requirements from a synthesis of domain analysis, interviews and tourist observations. We then proceed through four rounds of iterative prototyping; designing, building and testing at each stage. Rather than developing frameworks, as done in previous work \citep{Olsson2011}, we implement a full application. To empirically evaluate our system, we engage in tourism scenarios in a UK city (N = 20). Our participants judge the app to be usable, but find technical issues can disrupt AR. We finish by critically analysing the challenges of UCD.

\section{Related Work}

Vredenburg et al. (\citeyear{Vredenburg2002}) define user-centred design (UCD) as ``\textit{the active involvement of users for a clear understanding of user and task requirements, iterative design and evaluation, and a multi-disciplinary approach}''. The process involves participatory design where user feedback is solicited throughout development. As AR is often plagued by usability issues, UCD might deliver improvements.

We now reflect on existing AR tourism literature. Tokusho and Feiner (\citeyear{Tokusho2009}) developed an AR equivalent for Google StreetView. They found several usability challenges, including a limited field of vision. Although this tool also operated on an Android smartphone, its requirements were not informed by target users. Schinke et al. (\citeyear{Schinke2010}) suggested 3D arrows would contribute to usable navigation. While user studies indicated shapes were beneficial, their tests only included four POIs per screen.

Yovcheva et al. (\citeyear{Yovcheva2015}) created mock annotations, before evaluating them through a user study. Participants were found to value names and descriptions, and these findings will influence our early prototypes. While this work informed AR design, it did not produce an implemented system. Gabbard et al. (\citeyear{Gabbard1999}) constructed a methodology which combined task analysis, expert guidelines and user-centred evaluation. We apply a modified approach to AR tourism, extracting our requirements from participant observation and domain analysis.

With AR offering advantages over conventional apps, tourism tools have grown popular. Wikitude offers both AR browsers and development kits. Their app interfaces with Google Places, populating a locale with nearby attractions. However, it possesses usability issues, often obscuring the current location with POI annotations. ARNav offers similar functionality, even identifying mountains from their GPS position. Rather than crowding the screen with icons, the app presents a list of attractions to be selected. However, this approach also challenges usability, with POI selection being a cumbersome process. We learn from such works and move on to present our user-centred requirements gathering.

\section{Requirements Gathering}

\subsection{Domain Analysis}

Firstly, we designed our foundations on existing best practice. We surveyed Google Play apps using the search phrase of `augmented reality tourism'. We retrieved 192 applications, with 22 of these being AR, English-language and rated over 3/5. While we might have gained other lessons from a random sample, we sought best practice from the most-valued tools. We found overhead maps and icon annotations to be popular, with these features added to our requirements. We then surveyed user studies and usability guidelines. Olsson and Salo (\citeyear{Olsson2011}) found visual cluttering to be troublesome, and so we sought to minimise this. Yovcheva (\citeyear{Yovcheva2015}) advised prioritising details for nearby attractions; another wise suggestion. By following guidelines by Nielsen (\citeyear{Nielsen1994}), we trust our app will be well-informed.

\subsection{Tourist Observation}
\label{sec:threeb}

To inform both our requirements and user interviews, we conducted participant observation with real tourists. 
Six adults were selected in situ from the general public of Oxford (UK), a tourist destination. Our participants were predominantly travelling with their families and were recruited beside a local attraction. We observed these individuals from a 10-metre distance for 10 minutes, before soliciting their experiences. All participants had planned their trips to popular attractions. They appeared more encouraged by images than by the history of iconic buildings. Several expressed that while they wished to learn about the sites, they had no access to this information. This implies that apps with offline content could be of benefit. Since route planning, features and decision-making seemed most important, we constructed our interview questions around these topics. 

\subsection{User Interviews}
\label{sec:threec}

Our semi-structured interviews were key to gathering final requirements. For this, we recruited 14 overseas students from a local university. While these users were not tourists, they were not strongly acquainted with the local environment. The interviews were contextualised around a tourism scenario, with durations ranging from 30 to 45 minutes. Participants were asked 8 questions, with their responses manually noted. As most individuals reported using Google Maps to find attractions, a tourism app might be beneficial. 

Respondents next ranked app features in order of importance. The highest-rated functions were top attractions and local restaurants, with events deemed least important. Suggested features included nearby toilets and optimised routes, with these points added to our requirements. To assess decision-making, participants disclosed what influences their POI selection. We found distance and reputation were most important, with these opinions directly fed into our requirements.

\section{Iterative Development}

\textbf{Stage 1: Initial Mock-ups}. To inform our iterative development, we recruited 10 distinct overseas students. Tourist participation was impractical for a process which requires a consistent sample for several weeks. We first created low-fidelity mock-ups, enabling rapid refinement. We developed five annotations around three factors: size, clarity and detail. When soliciting user feedback, 80\% agreed that POI distance assists navigation. They also claimed ratings can help filter out undesirable premises. We therefore selected four annotation details: name, type, distance and rating.

\textbf{Stage 2: Map Prototypes}. Without an overhead projection, annotations can lose spatial relevance. Therefore, before we implemented AR, we prototyped our maps. We developed a skeleton Android app, populating attractions with Wikipedia data. In our participant feedback, users complained that POI navigation was cumbersome. Reacting to this, we refined interface transitions. As details were now revealed by an upward swipe, information could be browsed without changing screens. 

\begin{figure}[h!]
    \includegraphics[width=0.46\textwidth]{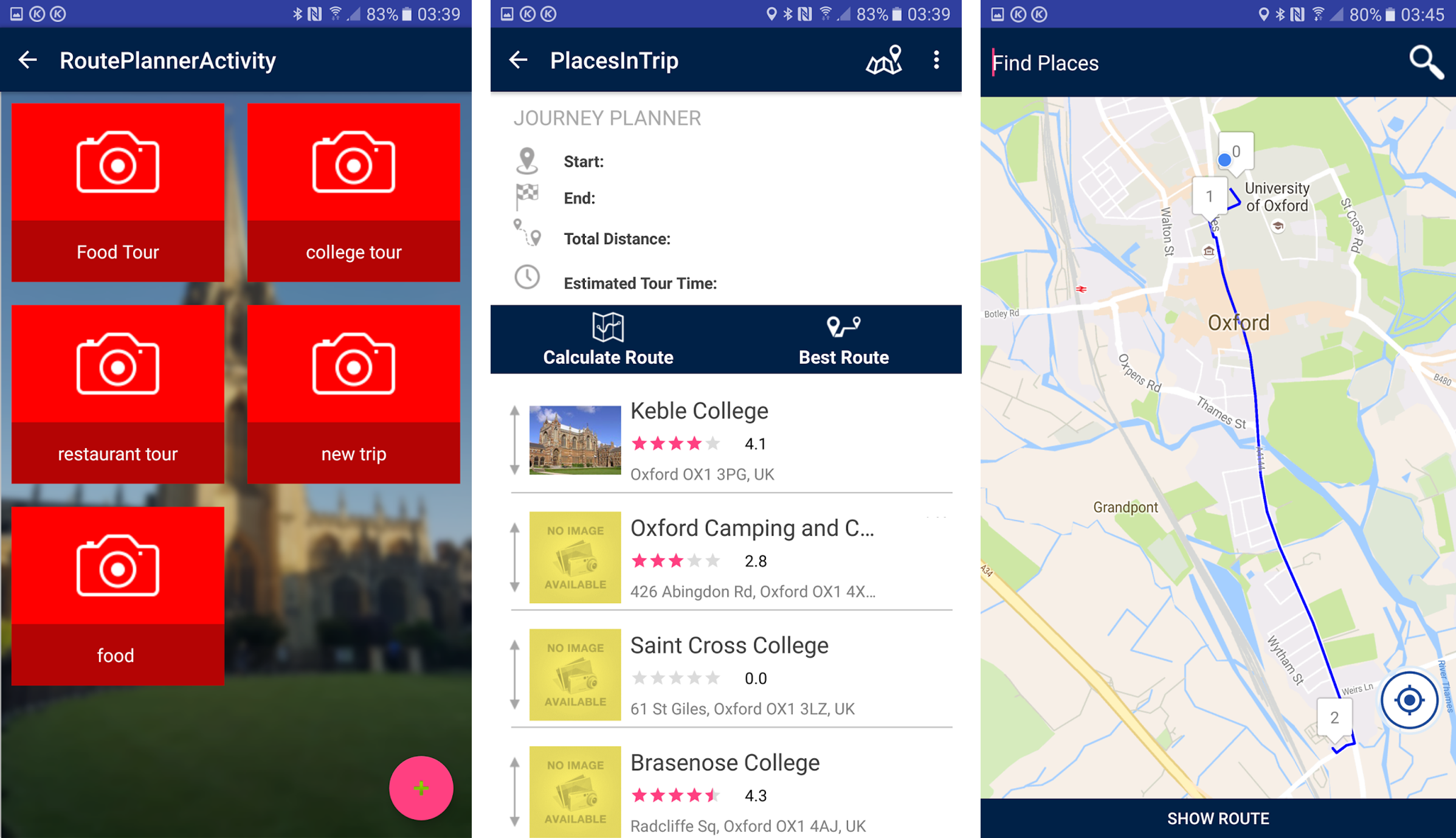}
    \centering
    \caption{Initial route planner prototype}
    \label{fig:route}
\end{figure}

\textbf{Stage 3: Route Planner Prototypes}. Through both our domain analysis and tourist observation, route planners appeared popular. In our enhanced prototype (Figure \ref{fig:route}), users created trips by selecting POIs. We designed icons for each type, with bars represented by a cocktail glass. 

To test the usability of our planner, we requested user feedback. Participants noted that once POIs were included, they could not be removed. They also disliked the set tour sequence, as tourists can arrive from different locations. Refining the app, we allowed POIs to be toggled through checkboxes. Start and end points also became adjustable, with routes automatically updated.

\begin{figure}[h!]
    \includegraphics[width=0.46\textwidth]{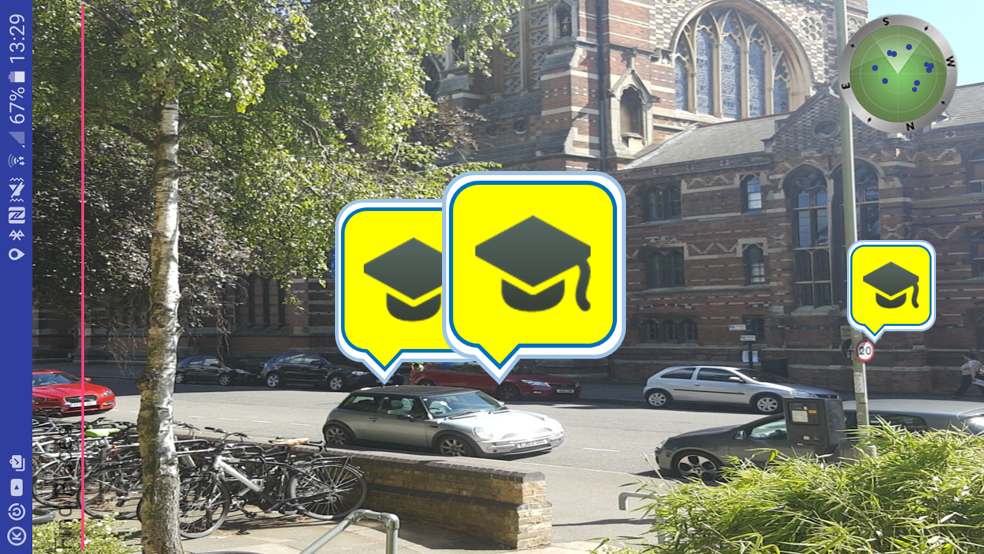}
    \centering
    \caption{Initial AR browser prototype}
    \label{fig:ar1}
\end{figure}

\textbf{Stage 4: AR Browser Prototypes}. We prototyped a basic browser (Figure \ref{fig:ar1}), including a search bar and icon overlays. Through comments, users suggested that annotation size should represent proximity. We also found that individuals only used landscape orientation for the AR browser. Reacting to these comments, we added auto-rotate functionality. To test the annotations, we replaced our icons with three alternatives: text, detailed text and images. 

Most participants preferred the `detailed text' annotations, as they provided more data. However, users claimed the verbosity led to screen cluttering. To account for this, we allowed the overlay size to be adjusted. With AR unused when the phone faces downwards, we configured the interface to automatically switch to the map.

\begin{figure}[h!]
    \includegraphics[width=0.46\textwidth]{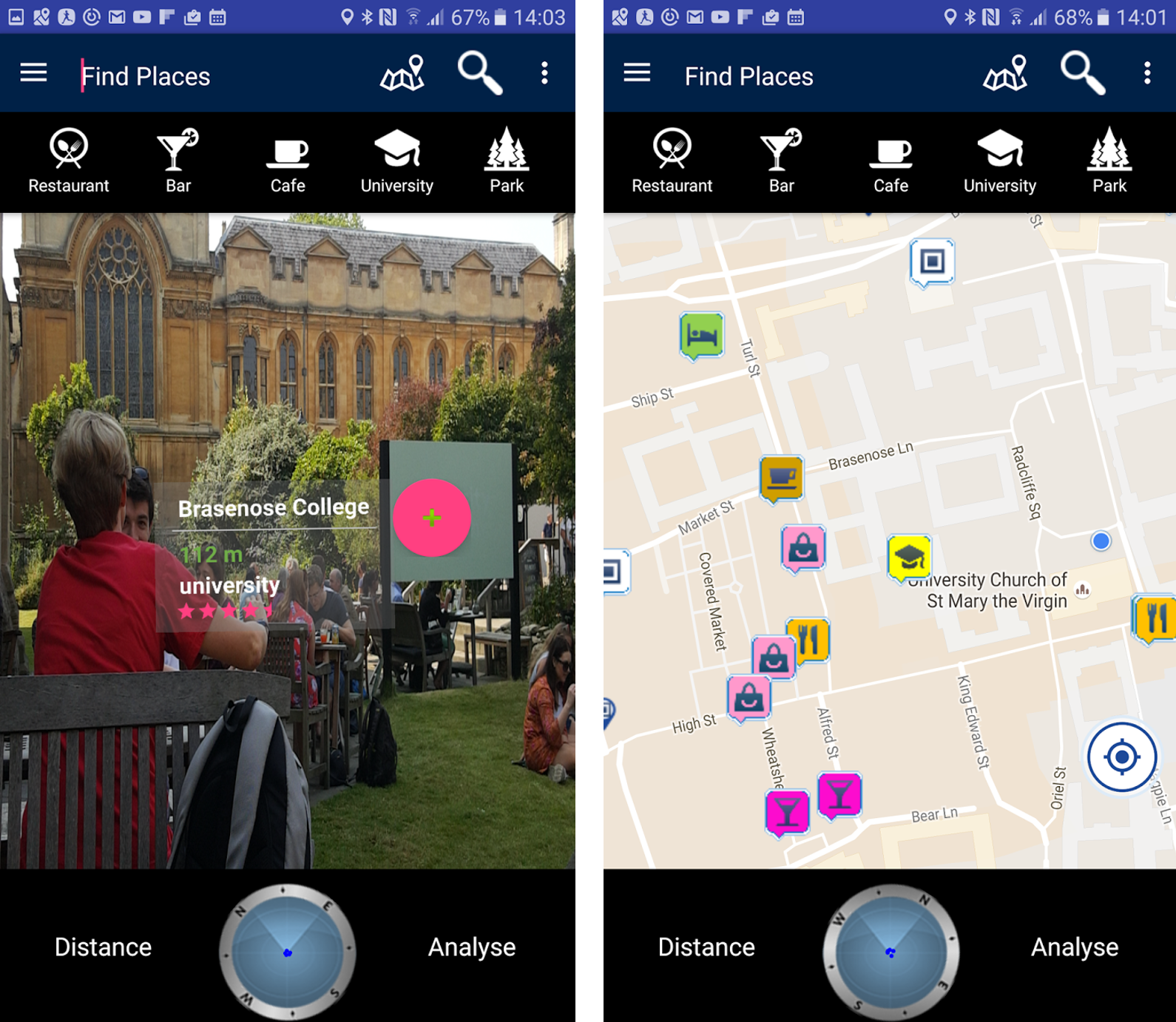}
    \centering
    \caption{Final AR tourism app}
    \label{fig:final}
\end{figure}

\textbf{Stage 5: Final Design}. Before completion, we added one additional feature: offline content. This functionality was requested in our observations, as many tourists lack Internet access. Through following a UCD approach, we believe our final \textit{ToARist} app (Figure \ref{fig:final}) to be well informed.

This tool has been tested and refined through an iterative process. However, to ascertain whether \textit{ToARist} is usable, we move on to evaluate the system in its entirety.

\section{Evaluation and Discussion}

Rather than using the same participants, we wished to explore the wider applicability of their opinions. To both validate our design and evaluate at a larger scale, we recruited 10 additional overseas students. We feel this approach was more feasible than inconveniencing 20 holidaymakers. To test the system in a real-world environment, we developed a 60-minute tourism activity. Users were observed navigating Oxford (UK) using \textit{ToARist}. They opened the browser at defined points, using the interface to explore their surroundings. After they finished the activity, they completed an exit questionnaire. 

\subsection{Findings and Discussion}

Most participants remarked that the app was simple and usable. Several even claimed they would be more likely to explore their area if they had the tool. Rather than concerning design, the most frequent complaint came from phone hardware. Several users encountered magnetic interference, which reduced the accuracy of POI annotations. While the icon overlays offered a high-level overview, many preferred the map unless in close proximity. Since our users regarded the app as usable, we believe this extols the benefits of user-centred design.

Reflecting on our findings, we found that tourist navigation can be grouped into three categories. Some users planned their trip in detail and visited attractions through the shortest path. Other individuals attended the POIs but navigated the city flexibly. Yet other users were exploratory, locating new attractions in situ. A successful AR tourism app must cater for the needs of all three groups.

While users appreciated our overlays, annotations often cluttered the screen. When POIs are distant, small icons could be used to advertise attractions. Since distance was displayed in the browser, users also used these details to locate POIs. As bearings are prone to magnetic interference, this data could prove helpful to disorientated tourists. The transition gesture proved popular, with it enabling quick navigation between the map and browser. This suggests tourists would use both tools to locate attractions. If users visualise distant POIs better on a map, this choice should not be obstructed. With screen size limited on phones, AR designers should further explore the role of gestures.

\subsection{User-Centred Design Challenges}

While we believe UCD offers benefits, we would like to highlight our challenges. Users often requested the perfect system: one that was functional, usable and attractive. Contradictory requests were frequent, most often originating from different individuals. As users express a wide range of opinions, there can be a temptation to design by committee. We were required to make several executive decisions, as a composite approach might have crippled usability. Development without best practice could produce an undesirable hybrid of subjective suggestions. There were also practical issues, with each round of development delayed by feedback. While no users withdrew from our process, enthusiasm waned as the study progressed.

\section{Conclusion and Future Work}

We developed an AR tourism app through an iterative process of user-centred design. Requirements were informed by ordinary users, achieved through a synthesis of domain analysis, tourist observations and interviews. We prototyped our application, feeding user opinions back into the design. The app was evaluated through a live scenario with 20 participants. Through performing real tourist exercises, our users found the AR browser to be usable. We finally reflected on our findings and the challenges of UCD.

Despite our contribution, we accept limitations to our work. Firstly, while our sample is not insignificant, it would have benefited from more participants. Secondly, whereas our users valued the app, their opinions were not made relative to other systems. We would therefore like to evaluate our tool against popular alternatives. Since earlier prototypes lacked AR functionality, we could compare their success to that of later versions. After considering feedback, we developed several suggestions for further work. Our participants praised the gesture which transitioned from the browser to the map. Future studies could explore the role of AR gestures and whether they can simplify cluttered interfaces. With the overhead map preferred for distant POIs, AR should attempt to enhance the experience. This could be achieved through 3D isometric projections, with the view updated based on smartphone sensors.

\bibliographystyle{apsr}
\bibliography{bib}

\end{document}